\documentclass[10pt,twocolumn,prl,aps,amssymb,amsmath,tightenlines]{revtex4}

\usepackage{dsfont}
\usepackage[dvips]{graphicx}
\begin{document}
%
%

%


\maketitle

\noindent \textbf{Comment on ``Typicality for Generalized
Microcanonical Ensemble''} \\

In a recent Letter, Reimann~\cite{reimann} has proposed an
interesting heuristic argument for the typicality of a particular
microcanonical ensemble. However, closer examination reveals that
the claimed results cannot be substantiated.

The main claim of Ref.~\cite{reimann} can be summarised as follows:
Provided that a probability distribution over the space of quantum
states is such that (i) it is uniformly distributed over the
relative phase variables and the amplitude variables are
independently distributed; and (ii) the associated density matrix
satisfies ${\rm tr}\rho^2\ll1$, then the quantity
$\sigma_{A_\psi}^2$ defined by
\[
\sigma_{A_\psi}^2 = {\mathbb E} \left[(A_\psi-{\bar A})^2\right]
\]
is small. Here ${\bar A}= {\mathbb E}[A]$ denotes the ensemble
average (unconditional expectation) of $A$, and $A_\psi=
\langle\psi|A|\psi\rangle$ denotes the expectation of $A$
conditional on the random pure state $|\psi\rangle$
(cf.~\cite{Khinchin}). In what follows I shall examine (a) the
plausibility of the two assumptions, (b) the validity of the claim
that $\sigma_{A_\psi}^2$ is small for generic observables, and (c)
the feasibility of the so-called typicality argument.

(a) First, consider the plausibility of the proposed equilibrium
states. The condition (i) on $\{\phi_n\}$ is a classical result that
can be established explicitly~\cite{brody}, whereas justification
for the independence of $\{\rho_n\}$ is missing~\cite{kotz}. As
regards condition (ii), one can show that the values of $\rho_n$ are
completely determined by the initial values of a commuting family of
observables that includes the Hamiltonian~\cite{brody}. Therefore,
\textit{a priori} one cannot decide whether this condition is
plausible without specific knowledge of the system under
consideration. Otherwise stated, for a closed and isolated system,
${\rm tr}\rho^2$ is a constant of motion, and thus one cannot decide
whether it should be large or small from the dynamics.

(b) As regards the magnitude of $\sigma_{A_\psi}^2$, in
Ref.~\cite{reimann} the author obtains the upper bound $\Delta_A^2\,
{\rm tr}\rho^2$, where $\Delta_A$ is the difference between the
largest and the smallest eigenvalues of $A$. Clearly, ${\rm tr}
\rho^2$ assumes the minimum value $N^{-1}$ if ${\bar\rho}_n= N^{-1}$
for all $n$. However, this is multiplied by $\Delta_A^2$, and if the
spectrum of $A$ grows linearly in the levels (like the energy
eigenvalues of a spin system or a harmonic oscillator), then we have
$\Delta_A^2\sim N^2$, which outweighs ${\rm tr}\rho^2\sim N^{-1}$.
Indeed, one can calculate $\sigma_{A_\psi}^2$ directly by use of the
measure introduced in Ref.~\cite{reimann} to establish that
$\sigma_{A_\psi}^2$ can be either very large or as small as zero,
depending on the observable $A$. Thus, the claim that
$\sigma_{A_\psi}^2$ is small for an ``arbitrary observable'' $A$ is
false.

(c) I conclude by remarking that the variance resulting from
measurements of $A$ is determined by the sum
\[
\sigma_A^2={\mathbb E}\left[(A_\psi-{\bar A})^2\right]+{\mathbb E}
[(A^2)_\psi-A_\psi^2]
\]
of the \textit{variance of the conditional expectation} and the
\textit{expectation of the conditional variance}, where $(A^2)_\psi=
\langle\psi|A^2|\psi\rangle$. Now the first term largely (but not
exclusively) reflects the statistical mixture of the state, and
vanishes, in particular, for pure states. The second term largely
(but not exclusively) reflects the pure quantum uncertainty of the
observable. If the first of the two terms is small for generic
observables, then it implies that the purity of the state is
correspondingly high, so that ${\rm tr}\rho^2\lesssim1$. Thus, any
attempt to establish the generality of the smallness of the first
term, starting from a highly mixed ensemble with ${\rm
tr}\rho^2\ll1$ as introduced in Ref.~\cite{reimann}, would be
futile. If the ensemble is highly mixed, then only under special
circumstances can this quantity be small. Therefore, starting with a
highly mixed ensemble satisfying the conditions (i) and (ii), one
cannot justify any microcanonical typicality argument, even if one
accepts the premise that small $\sigma_{A_\psi}^2$ implies
typicality.

\vspace{0.5cm}

\noindent Dorje~C.~Brody

Department of Mathematics

Imperial College London

London SW7 2BZ, United Kingdom \\




\end{document}